\documentclass[12pt]{article}
\usepackage{setspace}
\usepackage{amsmath}
\usepackage[letterpaper, margin=1in]{geometry}
\usepackage{bm}
\usepackage[linesnumbered,ruled,vlined]{algorithm2e}
\usepackage{dsfont}
\usepackage{graphicx}
\usepackage{hyperref}
\usepackage{float}

\usepackage{natbib}
\author{Adam J. Stone and John Paul Gosling\\
	{\it Department of Mathematical Sciences, Durham University, UK}}
\date{\today}
\title{H-AddiVortes: Heteroscedastic (Bayesian) Additive Voronoi Tessellations}

\begin{document}
	
	\maketitle
	\begin{abstract}
    This paper introduces the Heteroscedastic AddiVortes model, a Bayesian non-parametric regression framework that simultaneously models the conditional mean and variance of a response variable using adaptive Voronoi tessellations. By employing a sum-of-tessellations approach for the mean and a product-of-tessellations approach for the variance, the model provides a flexible and interpretable means to capture complex, predictor-dependent relationships and heteroscedastic patterns in data. This dual-layer representation enables precise inference, even in high-dimensional settings, while maintaining computational feasibility through efficient Markov Chain Monte Carlo (MCMC) sampling and conjugate prior structures. We illustrate the model's capability through both simulated and real-world datasets, demonstrating its ability to capture nuanced variance structures, provide reliable predictive uncertainty quantification, and highlight key predictors influencing both the mean response and its variability. Empirical results show that the Heteroscedastic AddiVortes model offers a substantial improvement in capturing distributional properties compared to both homoscedastic and heteroscedastic alternatives, making it a robust tool for complex regression problems in various applied settings.\newline
		\textbf{Total Words:} 4037
	\end{abstract}
	
	\doublespacing
	\section{Introduction}\label{sec:intro}
	
The challenge of developing advanced regression techniques to capture the relationships between high-dimensional predictors \(\bm{x}\) and a continuous response \(Y\) remains complex. Although notable approaches, including Gaussian process regression \citep{GpRegression}, random forests \citep{RandomForests}, gradient-boosting machines \citep{Boosting}, and deep neural networks for regression (DNNR) \citep{DNNR}, have been developed, many are limited to modeling only the conditional mean \(\mathds{E}[Y|\bm{x}]\) with a constant error variance assumption. Bayesian methods such as Bayesian Additive Regression Trees (BART) \citep{BARTpaper} offer robust tools for posterior inference and uncertainty quantification but often assume homoscedasticity, where the variance of \(Y\) given \(\bm{x}\) is independent of the predictors.

However, this assumption is unrealistic in many practical scenarios. Real-world data from many different industries such as medical data \citep{Zha_Heteroscedastic_MICCAI2024}, financial data \citep{Finance_Hetero} and climate data \citep{Climate_Hetero}, often exhibit heteroscedasticity, significantly influencing predictive performance and inferential accuracy. Approaches that address this complexity, such as heteroscedastic regression models, enable capturing relationships between predictors and the variability of the response.

Techniques for addressing heteroscedastic data include using weighted regression \citep{WeightedRegression}, which assigns different weights to data points based on the variance of their residuals. This helps to give less weight to observations with higher variance, leading to more efficient and unbiased estimates. However, determining the correct weights can be complex and may require prior knowledge about the nature of heteroscedasticity, and this method can be computationally intensive for large datasets. Transforming the dependent variable, such as taking the logarithm, can stabilize variance but complicate the interpretation of model coefficients. Using robust standard errors adjusts the standard errors of the regression coefficients to account for heteroscedasticity, providing more reliable statistical inferences, although it does not address the heteroscedasticity itself, which can still affect model performance.

In addition to these techniques, Gaussian Process (GP) regression variants provide sophisticated methods for handling heteroscedastic data. Warping \citep{Warping} involves applying nonlinear transformations to input data to better capture the underlying structure and heteroscedasticity. Heteroscedastic GP regression \citep{Hetero_GP} incorporates an input-dependent noise model that adapts the variance of the predictions based on the input space, making it particularly useful for datasets with varying noise levels. These GP variants offer flexible and powerful approaches to model complex, heteroscedastic relationships in data, but they also come with higher computational costs and require more expertise to implement effectively.

Recently, in \cite{HBART}, a nonparametric heteroscedastic elaboration of BART called HBART was proposed. In addition to the mean function being modeled with a sum of trees, each of which determines an additive contribution to the mean, the variance function is further modeled with a product of trees, each of which determines a multiplicative contribution to the variance. 

The present paper introduces the heteroscedastic AddiVortes model, a novel extension of Bayesian Additive Voronoi Tessellations (AddiVortes). The proposed methodology models both \(\mathds{E}[Y|\bm{x}]\) and \(\mathds{V}ar[Y|\bm{x}]\) through distinct multidimensional tessellations. The conditional mean is represented via a ``sum-of-tessellations'' ensemble that captures additive effects, while the conditional variance is modeled through a ``product-of-tessellations'' ensemble that multiplicatively defines the variance response surface. This dual ensemble approach ensures flexible, efficient modeling, benefiting from conditional conjugacy for computation and facilitating prior specification.

In high-dimensional contexts, this enhanced AddiVortes framework allows for the simultaneous inference of mean and variance structures, offering insights beyond conventional homoscedastic models. The flexibility of the model to capture complex interactions and heteroscedastic patterns makes it a valuable tool in diverse data analysis settings.

The remainder of the paper is organized as follows. In Section~\ref{Homo}, the homoscedastic AddiVortes model is described where the variance of the noise is assumed to be constant. Section~\ref{Hetero} builds upon the homoscedastic and explains the heteroscedastic AddiVortes model where the variance of the noise is dependent on the covariates. In Sections~\ref{simulated_data}, we illustrate the potential of heteroscedastic AddiVortes through simulated scenarios. In Section~\ref{Car_dataset} and Section~\ref{million_songs_dataset}, we compare the performance of heteroscedastic AddiVortes to heteroscedastic BART and the homoscedastic AddiVortes on real-world data using a range of metrics. Section~\ref{conclusion} concludes the paper with a discussion.

\section{The homoscedastic AddiVortes model}\label{Homo}

Many methods aim to model a conditional expectation function $f$ that relates the continuous variable $Y$ to some or all of $p$ (potential) covariates $\bm{x}=(x_1,\ldots,x_p)$, such that
\begin{equation*}
    Y=f(\bm{x})+\varepsilon, ~~~~~ \varepsilon \sim \mathcal{N}(0,\sigma^2),
\end{equation*}
under the assumption that the noise has zero mean and is homoscedastic.

In \cite{AddiVortes}, the AddiVortes model was introduced, which models the systematic part of this relationship using a sum of step-wise functions built upon Voronoi tessellations. More specifically, they approximated $f(\bm{x})=\mathds{E}(Y|\bm{x})$, the mean of $Y$ given $\bm{x}$, by the sum of many piecewise constant functions with boundaries defined by Voronoi tessellations.

The AddiVortes model consists of a predetermined fixed number, $m$, of tessellations, where $m$ takes the default value of 200 (a reasonable choice in most cases) or cross validation is applied. The $j^{\mathrm{th}}$ tessellation structure is denoted by $T_j$, that is, which covariates are dimensions in the tessellation and the coordinates of the centers of the cells. The corresponding output values for each cell are given by $\bm M_j=\{\mu_{1j},\ldots,\mu_{b_jj}\}$, where $b_j$ is the number of cells in tessellation $T_j$. The output value for a sample $\bm{x}$ of tessellation $T_j$,  is given by the function $g(\bm{x}|T_j,\bm M_j)$. An estimation of $\mathds{E}(Y|\bm{x})$ is given by the sum of all the outputs of the tessellations that $\bm{x}$ corresponds to, that is 

\begin{equation*}
    Y=\sum\limits_{j=1}^m g(\bm{x}|T_j,\bm M_j)+ \varepsilon, ~~~~~~ \varepsilon \sim \mathcal{N}(0,\sigma^2).
\end{equation*}

In single-dimension tessellations, the $\mu_{ij}$ represent main effects since $g(\bm{x}|T_j,\bm M_j)$ only depends on a single covariate but will represent interaction effects when the tessellations are multi-dimensional. Thus, AddiVortes can capture both the main effects and interaction effects in the model.

The aim of AddiVortes is to extract the posterior distribution of all unknown parameters in the sum-of-tessellations model,

\begin{equation}
    \pi((T_1, \bm M_1), \ldots, (T_m, \bm M_m), \sigma|\bm y_{\text{train}},\bm x_{\text{train}}).
\end{equation}

Thus, priors are specified for the parameters of the sum-of-tessellations model, specifically for $(T_1, \bm M_1),\ldots,(T_m, \bm M_m)$, and $\sigma$. These priors are strategically chosen to favor less complex configurations with fewer dimensions and cells. This regularization controls the influence of individual tessellation effects, preventing any one of them from becoming dominant. Without these regularization priors, complex tessellations with a large number of dimensions and centers would cause over-fitting and limit the advantages of the additive model in both approximation and computation. 

The specification of the regularization priors is simplified when independence restrictions are applied, such that,
\begin{equation}
    \begin{aligned}
        \pi\left(\left(T_1,\bm{M}_1\right),\ldots,\left(T_m,\bm{M}_m\right), \sigma\right) &=\prod\limits_{j=1}^m\left[\pi(T_j,\bm{M}_j) \right]\pi(\sigma) \\
        &=\prod\limits_{j=1}^m \left[\pi\left(\bm{M}_j|T_j\right)\pi(T_j)\right]\pi(\sigma)
    \end{aligned}
    \label{independence}
\end{equation}
and
\begin{equation}
    \pi(\bm{M}_j|T_j)=\prod\limits_{i=1}^{b_j} \pi\left(\mu_{ij}|T_j\right),
    \label{independence2}
\end{equation}
where $\mu_{ij} \in \bm{M}_j$.

For $\sigma$, a conjugate prior is employed: a natural choice for this prior is the inverse chi-square distribution $\sigma^2 \propto {\nu \lambda}/{\chi^2_\nu}$, where $\nu$ and $\lambda$ are hyperparameters. We handle the parameters $\mu_{ij}$ using a conjugate normal distribution $N(\mu_\mu, \sigma^2_\mu)$ similar to one used in \cite{BARTpaper}, which confers significant computational advantages as we can easily marginalize $\mu_{ij}$.

The prior for the tessellation structure $T_j$ is specified by multiple factors:
\begin{enumerate}
  \item the number of covariates considered in $T_j$, 
  \item the number of centers in $T_j$,
  \item the covariates that are included in $T_j$ and
  \item the coordinates of the centers.
\end{enumerate}
In this paper, we apply the same prior distributions for these variables as the ones considered in \cite{AddiVortes}.

A Gibbs sampler is used that involves $m$ successive draws of $(T_j, \bm M_j)$ conditionally on $(\{T_{j'}, \bm M_{j'}\}_{j \neq j'}, \sigma, y)$ for $j = 1, \ldots, m$, followed by a draw of $\sigma$ from its full conditional distribution. \cite{MCMCBackfitting} previously explored a similar application of the Gibbs sampler for posterior sampling in additive and generalized additive models, with $\sigma$ held fixed.

The draw of $\sigma$ from the full conditional can simply be achieved by sampling from an inverse gamma distribution. However, to sample from $(T_j, \bm M_j) | (\{T_{j'}, \bm M_{j'}\}_{j \neq j'}, \sigma, y)$, it is important to note that the conditional probability $\pi(T_j,\bm{M}_j|\{T_{j'} , \bm{M}_{j'}\}_{j \neq j'},\sigma, \bm{y})$ relies solely on $(\{T_{j'} , \bm{M}_{j'}\}_{j \neq j'}, \bm{y},\bm{X})$ through the expression
\begin{equation*}
\bm{R}_j=\bm y-\sum\limits_{k\neq j}g(\bm x|T_k,\bm{M}_k),
\end{equation*}
where $\bm{R}_j$ represents the $n$-vector of partial residuals derived from a fitting process that excludes the $j^\mathrm{th}$ tessellation. Now, $(T_j ,\bm{M}_j )~|~\bm{R}_j , \sigma$, is formally equivalent to the posterior of the single tessellation model $\bm{R}_j = g(x;T_j ,\bm{M}_j ) +\epsilon$ where $\bm{R}_j$ plays the role of the data $\bm{y}$.

We are able to carry out each draw in two successive steps as
\begin{equation}
\begin{aligned}
    &T_j |\bm{R}_j , \sigma, \\
    &\bm{M}_j |T_j , \bm{R}_j , \sigma.
\end{aligned}
\label{steps}
\end{equation}

The draw of $T_j$ in \eqref{steps} can be obtained using a Metropolis–Hastings (MH) algorithm. Six moves are suggested to propose a new tessellation based on the current tessellation, each with its associated proposal probability:
\begin{itemize}
  \item adding/removing a center (0.2 each),
  \item adding/removing a covariate (0.2 each),
  \item swapping a covariate (0.1) or
  \item changing the position of a center (0.1).
\end{itemize}

Subsequently, the draw of $ \bm M_j$ in Equation~\eqref{steps} involves independent draws of $\mu_{ij}$ from a normal distribution since we use a conjugate normal distribution.

By running the algorithm for a sufficient number of iterations after a suitable burn-in period, each iteration can be treated as an approximate, dependent sample from $\pi(f|y)$. To estimate $f(x)$, we can take the average or median of the posterior samples. A reasonable $(1 - \alpha)\%$ posterior interval for $f(x)$ is the interval between the upper and lower $\alpha/2$ quantiles and these uncertainty intervals behave sensibly, widening at $x$ values far from the data.

\section{Heteroscedastic AddiVortes}\label{Hetero}

In this section, we extend the AddiVortes model to accommodate data with non-constant variance. In many real-world situations, data do not adhere to the simple constant-variance assumption. The original AddiVortes model assumes a homoscedastic error structure, where the variance of the response \(Y\) remains constant regardless of predictor values. However, practical data frequently exhibit a covariate-dependent variance function. Our heteroscedastic AddiVortes approach jointly models the conditional mean function \(\mathds{E}[Y|\bm{x}] = f(\bm{x})\) and the conditional variance function \(\mathds{V}ar[Y|\bm{x}] = s^2(\bm{x})\), leading to a more realistic and flexible modeling framework. Specifically, we express:
\[
Y = f(\bm{x}) + \epsilon, \quad \epsilon \sim N(0, s^2(\bm{x})),
\]
where \(\bm{x} = (x_1, x_2, \ldots, x_d)\) is a \(d\)-dimensional vector of predictors. While we assume for simplicity that the same predictors influence both \(f(\bm{x})\) and \(s^2(\bm{x})\), this assumption is not strictly necessary.

Our methodology uses two distinct ensembles of Voronoi tessellations: an additive tessellation model for the mean function, denoted by \(f(\bm{x}) = \sum_{j=1}^m g(\bm{x}; T_j, M_j)\), and a multiplicative regression tessellation model for the variance component:
\[
s^2(\bm{x}) = \prod_{l=1}^{m'} h(\bm{x}; T'_l, M'_l).
\]
Here, \(g(\bm{x}; T_j, M_j)\) represents the output value of the \(j^\text{th}\) tessellation for the mean, while \(h(\bm{x}; T'_l, M'_l)\) represents the output value of the \(l^\text{th}\) tessellation for the variance.

In this formulation, \(T_j\) denotes the structure of the \(j^\text{th}\) tessellation for the mean, with \(M_j = \{\mu_{1j}, \ldots, \mu_{b_j,j}\}\) representing the \(b_j = |M_j|\) scalar cell output values. Similarly, \(T'_l\) denotes the structure of the \(l^\text{th}\) tessellation for the variance, and \(M'_l = \{s^2_{1l}, \ldots, s^2_{b'_ll}\}\) represents the \(b'_l = |M'_l|\) scalar cell output values. This approach allows \(f(\bm{x})\) to be modeled additively, capturing the effects through a sum of tessellations, while \(s^2(\bm{x})\) is modeled multiplicatively, capturing the scale through a product of tessellations.

	\subsection{Mean Model}

The mean is modeled in a manner nearly identical to the original AddiVortes, except that the variance of the error term is no longer assumed constant. Similar to the homoscedastic setting, we sample from \((T_j, \bm{M}_j) \mid \bm{R}_j,\sigma\), where 
\[
R_{ij} = y_i - \sum_{q \neq j} g(\bm{x}_i; T_q, \bm{M}_q).
\] 
Recall that this is formally equivalent to the posterior of the single tessellation model \(\bm{R}_j = g(\bm{x}; T_j, \bm{M}_j) + \epsilon\), where \(\bm{R}_j\) plays the role of the data \(\bm{y}\). Thus, when viewed as a function of \(\mu_{kj}\), the heteroscedastic AddiVortes likelihood in the \(k^\text{th}\) cell of the \(j^\text{th}\) mean tessellation is given by
\begin{equation}
    L(\mu_{kj} \mid \cdot) = \prod_{i=1}^{n_{kj}} \frac{1}{\sqrt{2 \pi s(\bm{x}_i)}} \exp \left( -\frac{(R_{ij} - \mu_{kj})^2}{2s^2(\bm{x}_i)} \right),
\end{equation}
where \(n_{kj}\) denotes the number of observations that map to the particular cell. As in the original AddiVortes model, we assume a conjugate prior distribution for the mean component, where \(\pi(\mu_{kj}) \sim N(0, \sigma_\mu^2)\) for all \(k,j \). The full conditional distribution for \(\mu_{kj}\) is given by
\begin{equation}
    \pi(\mu_{kj} \mid \cdot) \sim N \left( \frac{\sum_{i=1}^{n} \frac{R_{ij}}{s^2(\bm{x}_i)}}{\frac{1}{\sigma_\mu^2} + \sum_{i=1}^{n} \frac{1}{s^2(\bm{x}_i)}}, \frac{1}{\frac{1}{\sigma_\mu^2} + \sum_{i=1}^{n} \frac{1}{s^2(\bm{x}_i)}} \right).
\end{equation}

The integrated likelihood, ignoring terms that cancel in the acceptance probability, is given by:
\begin{equation}
    \int L(\mu_{kj} \mid \cdot) \pi(\mu_{kj}) \, d\mu_{kj} \propto \left( \sigma_\mu^2 \sum_{i=1}^{n} \frac{1}{s^2(\bm{x}_i)} + 1 \right)^{-1/2} \exp \left( -\frac{\sigma_\mu^2 \left( \sum_{i=1}^{n} \frac{R_{ij}}{s^2(\bm{x}_i)} \right)^2}{2 \left( \sigma_\mu^2 \sum_{i=1}^{n} \frac{1}{s^2(\bm{x}_i)} + 1 \right)} \right).
\end{equation}

These forms are nearly identical to those in the original AddiVortes model, with the main difference being the replacement of the scalar variance \(\sigma^2\) with the vector variance \(s^2(\bm{x}_i)\) for \(i = 1, \ldots, n\).

	\subsection{Variance Model}\label{Variance_Model}

We assume that
\[\bm Y \sim \mathcal{N}(f(\bm x),s^2(\bm x))\]
where $s^2(\bm x)= \prod_{l=1}^{m'} h(\bm{x}; T'_l, M'_l)$. Therefore, for any \(i = 1, \ldots, n\) we have  
\[ \mathds{P}(\bm Y|f(\bm x),s(\bm x_i)) \sim  \frac{1}{\sqrt{2 \pi s(\bm x_i)}} \exp \left( -\left (\frac{y_i-f(\bm x_i))^2}{2s^2(\bm x_i)} \right) \right)
\sim  \frac{1}{\sqrt{2 \pi s_{kl}}} \exp \left( -\frac{(\frac{y_i-f(\bm x_i)}{s^2_{-l}(\bm x_i)})^2}{2s_{kl}} \right)\]
where $s_{kl}$ is the output value of tessellation $l$ for $\bm x_i$ for the variance model.
    
	The Heteroscedastic AddiVortes likelihood function for the variance component in the \( k^\text{th} \) cell of the \( l^\text{th} \) variance tessellation is given by:
	
	\begin{equation}
		L(s^2_{lk} | \cdot) = \prod\limits_{i=1}^{n_{kl}} \frac{1}{\sqrt{2 \pi s_{lk}}} \exp \left( -\frac{e_i^2}{2s^2_{lk}} \right),
	\end{equation}
	
	where \( n_{kl} \) represents the number of observations in the cell, and \( e_i^2 = \frac{\left( y_i - \sum_{j=1}^{m} g(\bm x_i; T_j, \bm M_j) \right)^2 }{ s_{-l}^2(\bm x_i)} \), where, \( s_{-l}^2(\bm x_i) = \prod\limits_{q \neq l} h(\bm x_i; T'_q, M'_q) \). A conjugate prior distribution is specified for the variance component, \( s^2_{lk} \sim \chi^{-2}(\nu', \lambda') \), for all \( l \) and \( k \). The full conditional distribution for \( s^2_{lk} \) then becomes:
	
	\begin{equation}
		s^2_{lk} | \cdot \sim \chi^{-2} \left( \nu' + n, \frac{\nu' \lambda' + \sum_{i=1}^{n} e_i^2}{\nu' + n} \right).
        \label{s_sample}
	\end{equation}
	
	This allows for efficient updates of the cells in the \( m' \) variance component tessellations within the product decomposition using Gibbs sampling.
	
	The integrated likelihood for \( s^2_{lk} \) is also available in closed form as:
	
	\begin{equation}
		\int L(s^2_{lk} | \cdot) \pi(s^2_{lk}) ds^2_{lk} = \frac{\Gamma \left( \frac{\nu' + n}{2} \right) \left( \frac{\nu' \lambda'^2}{2} \right)^{\nu'/2}}{ (2 \pi)^{n/2} \left( \nu' \lambda' + \sum_{i=1}^{n} e_i^2 \right)^{\frac{\nu' + n}{2}}}.
        \label{s_likelihood}
	\end{equation}
	
	This form depends on the data through the sufficient statistic \( \sum_{i=1}^{n} e_i^2 \), enabling the use of Metropolis–Hastings steps to explore the structure of the variance tessellations \( T'_1, \dots, T'_{m'} \) in a manner analogous to the mean model tessellations. That is, the conjugate inverse chi-squared prior leads to a Gibbs step drawing from an inverse chi-squared full conditional when updating the components of \( M' \) using \eqref{s_sample}. Sampling the tessellation structure \( T'_j \) is performed via Metropolis–Hastings steps using the marginal likelihood given in \eqref{s_likelihood}. This is made possible as the integrated likelihood is analytically tractable with the specified heteroscedastic model.
	
	The interpretation of this model follows from the mean being factored into a sum of weakly informative components (as usual in AddiVortes), while the variance is factored into the product of weakly informative components. 
	
	Note that the number of variance component tessellations, \( m' \), need not equal the number of mean component tessellations, \( m \). A default value that has worked well in the examples explored is \( m' = 40 \). Since the tessellations making up the mean model are different from those making up the variance model, the number of cells in the \( l^\text{th} \) variance component tessellation is unrelated to the number of cells in the \( j^\text{th} \) mean component tessellation. This means, for instance, that the complexity of the variance function may differ from that of the mean function, and the predictors that are important for the variance function may also differ from those that are important for the mean.
	

\subsection{Calibrating the Variance Prior}
	
This section demonstrates a straightforward strategy for selecting the prior parameters that yield reasonable and practical results.

As outlined in Section \ref{Variance_Model}, the prior for the variance is given by $s(\bm{x})^2 \sim \prod_{l=1}^{m'} s_l^2$, with $ s_l^2 \sim \chi^{-2}(\nu', \lambda'),$ where the \(s_l^2\) are independent and identically distributed. To determine the prior parameters \((\nu', \lambda')\) for the variance components, we begin by considering the prior for constant variance in the context of the homoscedastic AddiVortes model $Y = f(x) + \varepsilon, \quad \varepsilon \sim N(0, \sigma^2)$, where $\sigma^2 \sim \chi^{-2}(\nu, \lambda)$. \cite{AddiVortes} proposed strategies for selecting \((\nu, \lambda)\) in this scenario. To ensure consistency between the priors in the heteroscedastic and homoscedastic models, we match the prior means of the variance components.
 We have:
	
	\[
	\mathds{E}[\sigma^2] = \frac{\nu \lambda}{\nu - 2}, \quad \text{and} \quad E\left[s(x)^2\right] = \prod\limits_{l=1}^{m'} E\left[s_l^2\right] = \lambda^{m'} \left( \frac{\nu'}{\nu' - 2} \right)^{m'}.
	\]
	
	By equating these means, we ensure compatibility between the heteroscedastic and homoscedastic priors. To achieve this, we separately match the “$\lambda$ component” and the “$\nu$ component”, leading to the following expressions for the parameters of the heteroscedastic prior,
	
	\[
	\lambda' = \lambda^{\frac{1}{m'}}, \quad \nu' = \frac{2}{1 - \left(1 - \frac{2}{\nu}\right)^{\frac{1}{m'}}}.
	\]

\subsection{Model Evaluation Metrics}

Assessing model performance requires more than evaluating predictive accuracy alone, particularly when modeling both the mean and variance of the response. Traditional metrics such as the Root Mean Squared Error (RMSE) offer a straightforward measure of prediction error but may overlook crucial aspects of distributional fit. For models that aim to capture the full predictive distribution, it is equally important to evaluate properties such as interval coverage and the precision of uncertainty estimates.

To address these considerations, we introduce graphical diagnostics and distributional metrics. The \textit{H-evidence plot} is a graph of $\hat s(\bm{x}_i)$ verses the estimated posterior intervals of $\hat s(\bm{x}_i)$ for $i \in \{1,\ldots,n\}$, where $n$ is the number of observations. This plot offers an intuitive way to detect patterns in variance that depend on the predictors.

highlighting areas where heteroscedasticity is evident.

Another key diagnostic is the \textit{predictive quantile-quantile (QQ) plot}, which extends classical QQ plots to compare the empirical quantiles of observed data against those predicted by the model's full predictive distribution \(Y \mid \bm{x}\). If the model captures the data distribution well, the plotted points should approximate a straight line. Deviations from linearity indicate potential misfits, such as underestimating or overestimating variability.

In addition to graphical tools, we employ the \textit{e-statistic}, which is a measure designed to quantify the similarity between two distributions \( F_1 \) and \( F_2 \), represented by their respective random samples \(\{U_i\}\) and \(\{V_j\}\). This statistic provides a single scalar summary that captures both location and scale discrepancies. It is defined as:

\begin{equation}
e = \frac{2}{n_1 n_2} \sum_{i=1}^{n_1} \sum_{j=1}^{n_2} \|U_i - V_j\| - \frac{1}{n_1^2} \sum_{i=1}^{n_1} \sum_{j=1}^{n_1} \|U_i - U_j\| - \frac{1}{n_2^2} \sum_{i=1}^{n_2} \sum_{j=1}^{n_2} \|V_i - V_j\|,
\end{equation}

where \( \|\cdot\| \) denotes the Euclidean norm, and \( n_1 \) and \( n_2 \) are the sample sizes of \( \{U_i\} \) and \( \{V_j\} \), respectively.

The \textit{e-statistic} works by comparing the pairwise distances among points in the two sample sets. The first term in the equation calculates the average pairwise distance between samples in \( F_1 \) and \( F_2 \):

\[
\frac{2}{n_1 n_2} \sum_{i=1}^{n_1} \sum_{j=1}^{n_2} \|U_i - V_j\|.
\]
This term quantifies how far apart the two sets of samples are in Euclidean space. If the two distributions are similar, the distances between samples from \( F_1 \) and \( F_2 \) will tend to be small.

The second and third terms measure the pairwise distances within each sample set. Specifically,

\[
\frac{1}{n_1^2} \sum_{i=1}^{n_1} \sum_{j=1}^{n_1} \|U_i - U_j\| \quad \text{and} \quad \frac{1}{n_2^2} \sum_{i=1}^{n_2} \sum_{j=1}^{n_2} \|V_i - V_j\|.
\]
These terms reflect the spread or dispersion within the two distributions. By subtracting the internal distances from the between-sample distances, the e-statistic provides a measure of the dissimilarity between \( F_1 \) and \( F_2 \). A smaller value of \( e \) indicates closer agreement between the two distributions, while a larger value reflects greater discrepancy.

When comparing the \textit{uniform distribution} to the \textit{quantiles} of the sampled posterior distributions, the e-statistic can evaluate how closely the quantiles align with the uniform distribution. In this case, the uniform distribution on \([0,1]\) serves as the reference, representing the ideal case where data points are evenly spread across the interval. The quantiles, derived from the empirical distribution of the observed data, form the second set of samples.

The e-statistic serves as a rigorous measure for comparing two distributions. When applied to uniform distributions and empirical quantiles, it provides a clear assessment of how evenly the quantiles are spread, making it particularly useful for evaluating model calibration and fit.

For hyperparameter tuning, such as selecting the smoothing parameter \(\kappa\) in the prior mean model, cross-validation using the e-statistic ensures that both first- and second-moment properties are matched. This approach avoids relying solely on RMSE, which can be insufficient when modeling heteroscedasticity.

Overall, this combination of graphical diagnostics and quantitative metrics provides a comprehensive framework for evaluating and comparing the performance of the heteroscedastic AddiVortes model, ensuring that both predictive accuracy and uncertainty estimation are thoroughly assessed.

\section{Illustrating Heteroscedastic AddiVortes on simulated data}\label{simulated_data}
	
We use simulated data to evaluate the performance of AddiVortes against known values and consider a similar set-up to the one originally explored in \cite{Friedman}. We generate data by simulating the values of $\bm{x} = (x_1, x_2, \ldots, x_d)$, with $x_1, x_2, \ldots, x_d$ being independently drawn from the standard uniform distribution and the response variable $Y$ given by
\begin{equation}
		Y = f(\bm{x}) + \epsilon = 10 \sin(\pi x_1 x_2) + 20(x_3 - 0.5)^2 + 10x_4 + 5x_5 + \epsilon, \label{eq:Freiman}
	\end{equation}
	where $\epsilon \sim \mathcal{N}(0, s^2(\bm x))$. Notably, $Y$ is solely dependent on $x_1, \ldots, x_5$, so the predictors $x_6, \ldots, x_d$ are inactive variables. The introduction of these extraneous variables injects a layer of complexity. In our simulated experiments, we let $d=10$, that is there are 10 covariates with only 5 active covariates.
	
	To evaluate the predictive ability of the heteroscedastic AddiVortes model of the variance component $s^2(\bm x)$, we consider three cases,
	\begin{enumerate} 
        \item $s(\bm x)=1$,
		\item $s(\bm x)=5x_1+2x_2$,
		\item $s(\bm x)=\frac{1}{2}(10\sin(\pi x_1x_2)+20(x_3-0.5)^2+20x_4+5x_5)$.
	\end{enumerate}

The first case corresponds to the homoscedastic scenario, where the noise variance remains constant across the data. In contrast, the second and third cases involve heteroscedasticity, with non-constant noise variance. In the second case, the variance follows a simple linear relationship, while in the third case, it is determined by a more complex functional form.

\begin{figure}[H]
        \centering
        \includegraphics{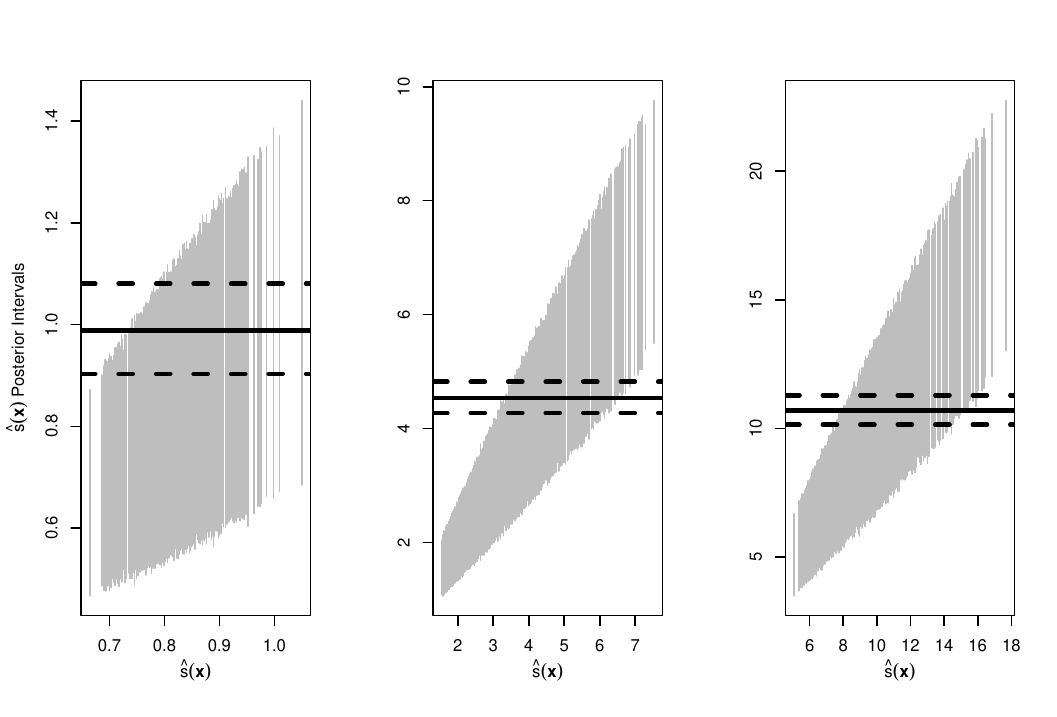}
        \caption{H-evidence plot for Heteroscedastic AddiVortes for case 1 (left), case 2 (middle) and case 3 (right). The solid horizontal line represents the estimate of $\sigma$ obtained from the homoscedastic AddiVortes fit, with the posterior 90\% interval depicted as dashed lines.}
        \label{fig:H_evidence_plot_simulated_data}
    \end{figure}

Figure \ref{fig:H_evidence_plot_simulated_data} displays the H-evidence plot for each of the cases. Observations are sorted based on the values of \(\hat{s}(\bm{x})\), with \(\hat{s}(\bm{x})\) plotted along the horizontal axis and the posterior intervals for \(s(\bm{x})\) shown on the vertical axis. A solid horizontal line is drawn at the estimate of \(\sigma\) obtained from the homoscedastic version of AddiVortes. In the homoscedastic case,  all the intervals overlap suggesting that the data is indeed not heteroscedastic. However, for the other cases, the separation of the posterior intervals from the horizontal line clearly indicates that the detection of heteroscedasticity is ``significant''. This plot allows us to assess the practical significance of deviations from constant variance and has proven useful across a range of applications.

\begin{figure}[H]
    \centering
    \includegraphics{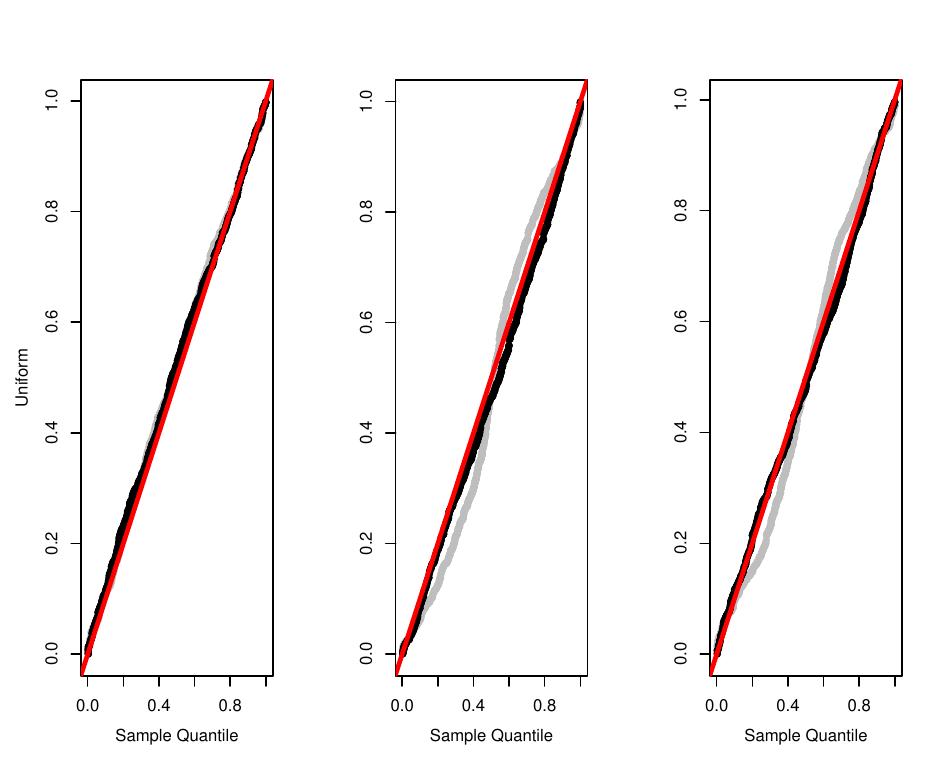}
    \caption{Predictive qq-plot for Heteroscedastic AddiVortes for case 1 (left), case 2 (middle) and case 3 (right). The solid horizontal line represents the estimate of $\sigma$ obtained from the homoscedastic AddiVortes fit, with the posterior 90\% interval depicted as dashed lines.}
    \label{fig:energy_plots_simulated}
\end{figure}

Figure \ref{fig:energy_plots_simulated} assesses the model fit using qq-plots derived from the predictive distribution generated by our model. For each \(i\), we obtain samples from \(p(y \mid \bm{x}_i)\) and compute the percentile of the observed \(y_i\) within these samples. If the model accurately captures the data-generating process, these percentiles should resemble draws from a uniform distribution over \((0, 1)\). The qq-plot is used to compare these percentiles against uniform draws. The black lines represent the qq-plots for the heteroscedastic AddiVortes whereas the gray lines represent the homoscedastic AddiVortes model.

In the first case, the lines are very similar since the homoscedastic model is able to capture the variance in the noise. However, in the other cases, the Heteroscedastic AddiVortes has a close alignment of the predictive percentiles with the uniform distribution suggesting an excellent model fit. The homoscedastic model's shortcomings are evident, despite both models exhibiting similar RMSE values (0.77 for the heteroscedastic AddiVortes and 0.76 for the AddiVortes model). In these models, the heteroscedastic AddiVortes model achieves an e-statistic score of 0.54, while the AddiVortes model scores 3.50. Although the RMSE values suggest that both models have comparably good mean prediction performance, the e-statistic and graphical diagnostics clearly demonstrate that the heteroscedastic AddiVortes model provides a substantially better overall fit.

\begin{figure}[H]
    \centering
    \includegraphics{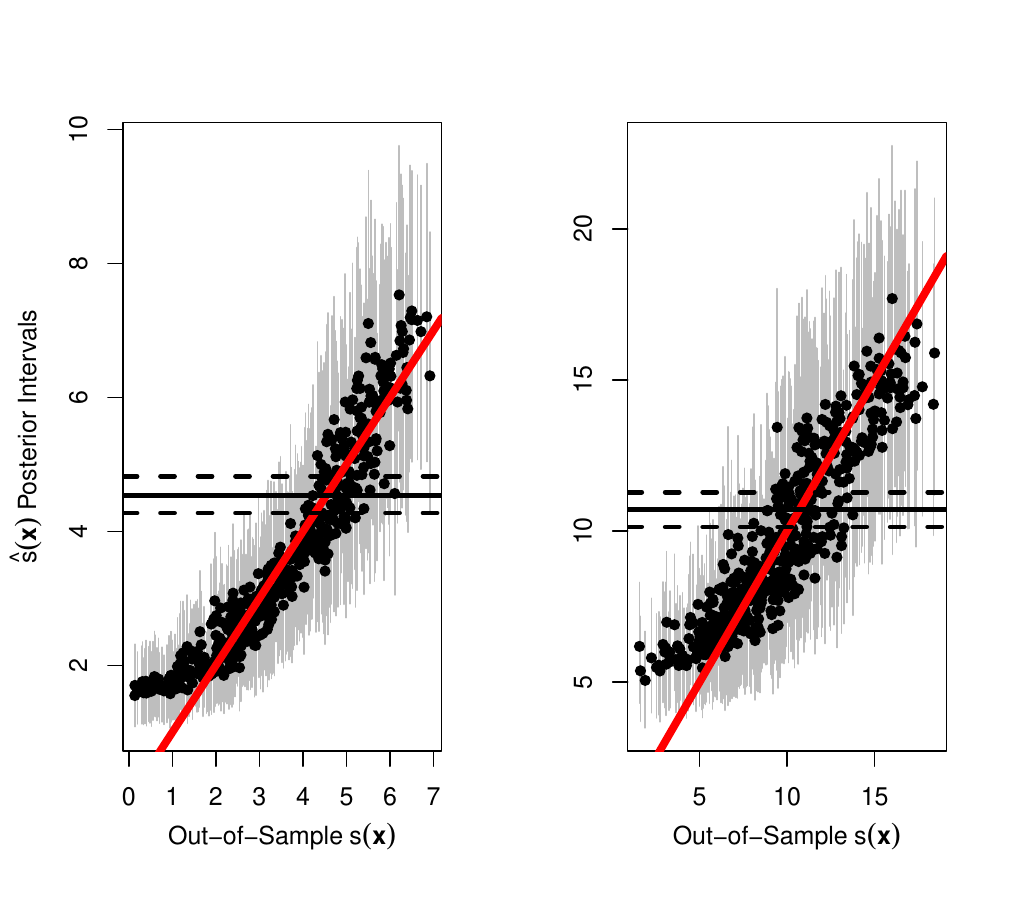}
    \caption{Comparison of true variance \(s(x)\) and predicted variance \(\hat{s}(x)\) when using the heteroscedastic AddiVortes model for case 2 (left) and case 3 (right).}
    \label{fig:s_vs_s_simulated_data}
\end{figure}

Since the data in this study is simulated, the true variance function \(s(\bm{x})\) is known for the testing data. Figure \ref{fig:s_vs_s_simulated_data} compares the actual variance \(s(\bm{x})\) to the predicted variance \(\hat{s}(\bm{x})\). The Heteroscedastic AddiVortes model demonstrates excellent predictive accuracy for the variance in both cases, despite being trained on a limited number of observations. Notably, the second case exhibits larger confidence intervals, reflecting increased uncertainty in the predictions, while in the first case, as the variance decreases, the model shows a slight tendency to overestimate the variance. This behavior underscores the model’s flexibility and its ability to capture nuanced variance structures effectively.

\begin{figure}[H]
    \centering
    \includegraphics{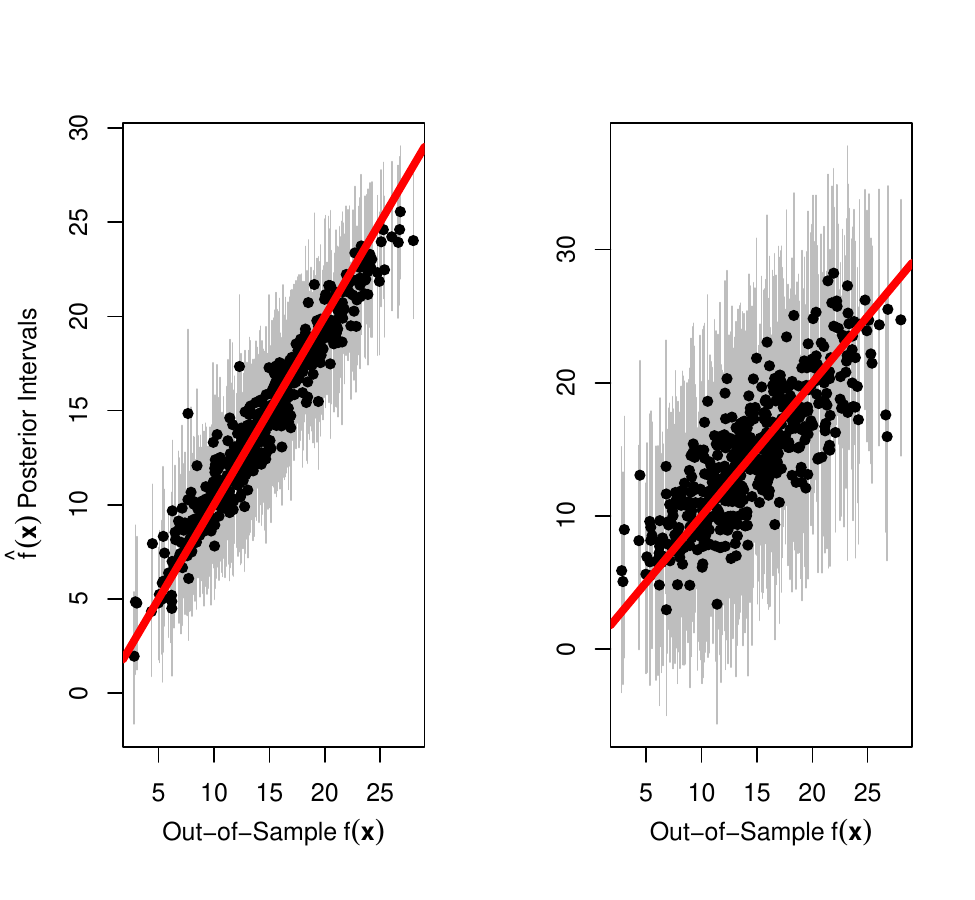}
    \caption{Comparison of the true mean function \(f(x)\) and predicted mean function \(\hat{f}(x)\) when using the heteroscedastic AddiVortes model for case 2 (left) and case 3 (right).}
    \label{fig:f_vs_f_simulated}
\end{figure}

Furthermore, figure \ref{fig:f_vs_f_simulated} shows the relationship between the true mean function \(f(\bm{x})\) and the predicted mean function \(\hat{f}(\bm{x})\). The close alignment of the predicted values to the actual values indicates that the Heteroscedastic AddiVortes model effectively captures the mean response as well, demonstrating strong predictive capabilities for both the mean and variance components of the data.

\section{Cars Dataset}\label{Car_dataset}

The heteroscedastic AddiVortes model was applied to a dataset of used car sales, featuring 1,000 observations collected from 1994 to 2013. This dataset, which spans the 2007–2008 financial crisis and subsequent recovery, includes both continuous predictors—mileage (in miles) and year of manufacture—and four categorical predictors: trim, color, displacement, and ownership status. After encoding the categorical variables into binary dummies, the dataset contained 14 predictors.

The goal was to explore the relationship between car prices and the predictors, particularly focusing on potential heteroscedasticity in price variability. The heteroscedastic AddiVortes model captured changes in variance that could not be explained by mean trends alone. For example, while higher mileage predictably led to lower prices, the variance of prices also showed complex dependence on mileage and year. Moreover, trim level, displacement, and ownership status emerged as key factors influencing the mean and the spread of car prices.

To tune the heteroscedastic AddiVortes model, cross-validation was performed using the e-statistic, ensuring that both mean and variance structures were accurately captured. This approach revealed that optimal settings for the hyperparameter \(\kappa\) differ for the heteroscedastic model compared to the standard AddiVortes model, with the former favoring more smoothing to account for variance heterogeneity. The resulting analysis provided a nuanced understanding of the predictors driving both the expected value and the variability of car prices, highlighting heteroscedastic AddiVortes' ability to uncover meaningful, data-driven insights in complex regression problems.

\begin{figure}[H]
        \centering
        \includegraphics{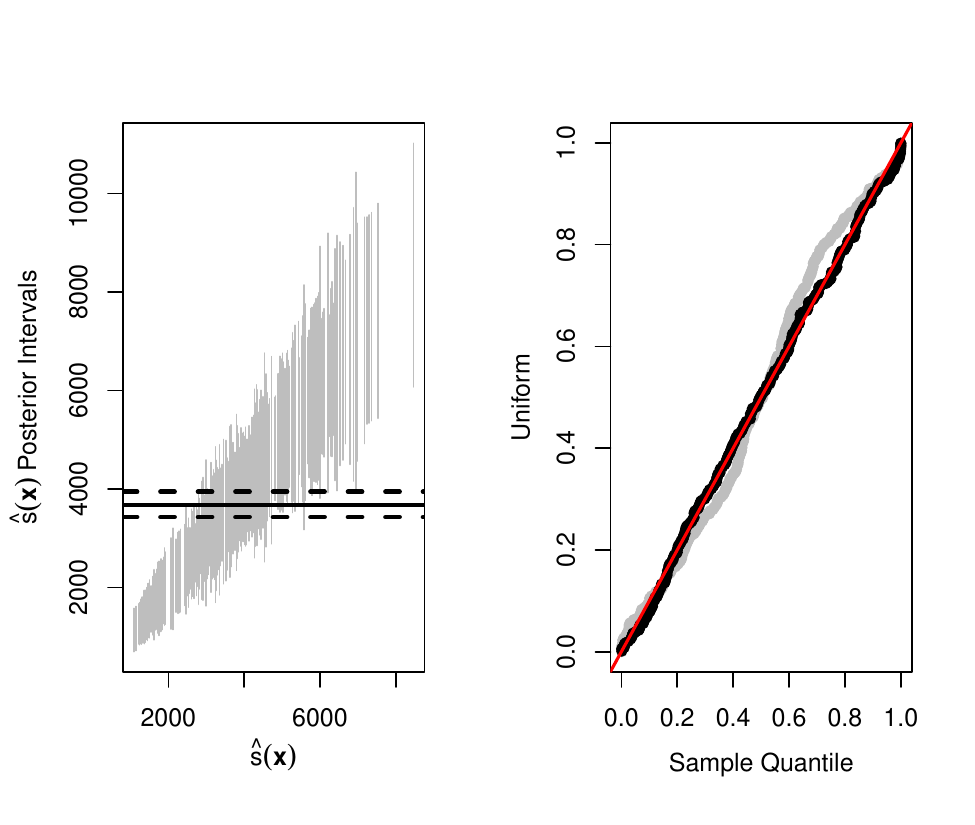}
        \caption{H-evidence plot (left) for Predictive qq-plot (right) for Heteroscedastic AddiVortes.}
        \label{fig:H-evidence_plot_car}
    \end{figure}

The corresponding predictive out-of-sample qq-plots for the homoscedastic model trained on 80\% of the full dataset with $\kappa = 0.75$ and the heteroscedastic model with $\kappa = 1.5$ are shown in Figure \ref{fig:H-evidence_plot_car}. This serves as an empirical graphical validation for the higher level of prior smoothing suggested by the cross-validated tuning of $\kappa$. The qq-plot for the heteroscedastic model is much closer to a straight line and dramatically better than the qq-plot for the homoscedastic version. 

The H-evidence plot in Figure \ref{fig:H-evidence_plot_car} serves as an additional check for evidence of heteroscedasticity: clearly, the posterior 90\% credible intervals do not cover the estimate of standard deviation from the homoscedastic model for a majority of the data. Even though there is considerable uncertainty about $s(x)$ at the higher levels, we have strong evidence that $s(\bm x) > 7000$ at the higher levels and $s(\bm x) < 2500$ at the lower levels. These differences are practically significant. 

\begin{figure}[H]
        \centering
        \includegraphics{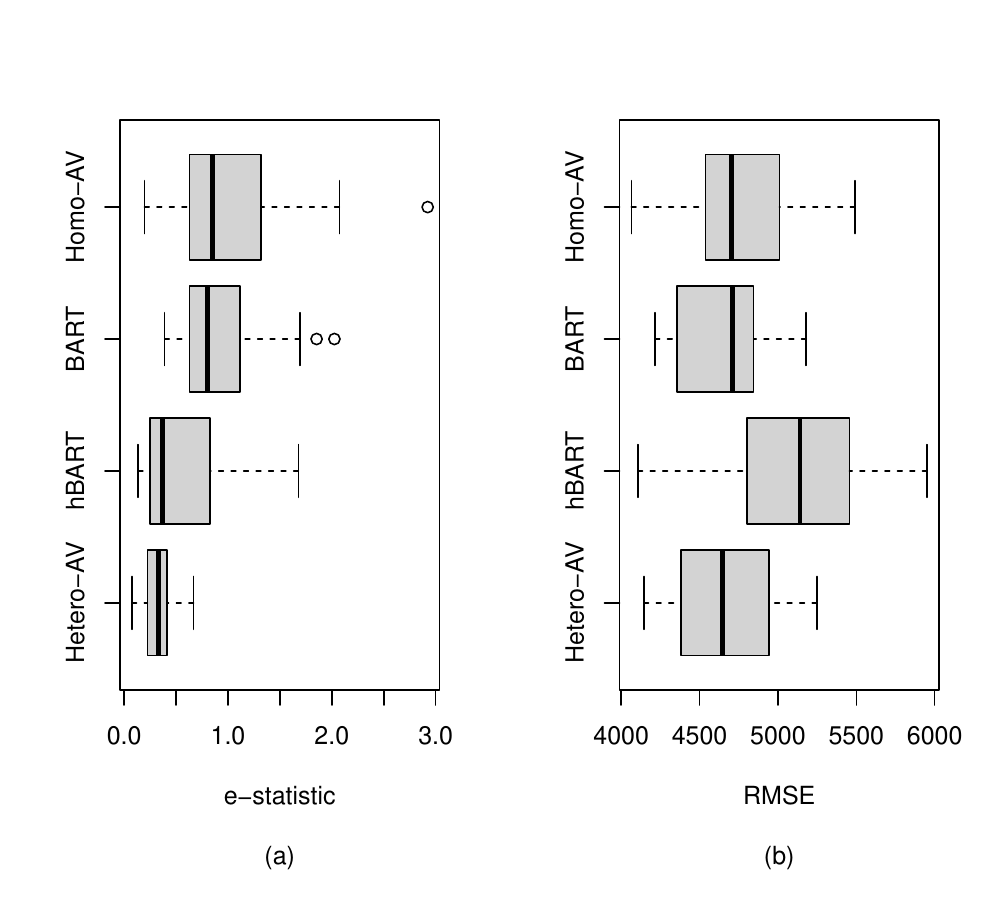}
        \caption{Box plot of e-statistic (left) and RMSE values (left) for all competing methods of the 20 train/test splits of the car dataset.}
        \label{fig:e-stat_RMSE_Boxplot_cars}
    \end{figure}

Now we compare the performance of heteroscedastic AddiVortes to HBART (\cite{HBART}, implement using \texttt{hbart} package), BART (\cite{BARTpaper}, implement using BayesTree package) and the homoscedastic AddiVortes model. To compare the performance of heteroscedastic AddiVortes to homoscedastic AddiVortes and the equivalent BART models we used cross-validation on the e-statistic using values in Table \ref{tab:CVvalues}, with all other hyperparameters taking their default value. We then took 20 random train/test samples, 80\%/20\% split, box plots of the RMSE values for each method across test/train splits are depicted in Figure~\ref{fig:e-stat_RMSE_Boxplot_cars} and the (50\%, 75\%) RRMSE quantiles are provided in Table~\ref{tab:quantiles}.

\begin{table}[ht]
  \centering
  \begin{tabular}{lll}
    \hline
    \textbf{Method} & \textbf{Parameters} & \textbf{Values considered} \\
    \hline
    Homos-AddiVortes 
        & Sigma prior: $(\nu,q)$ & (3, 0.90), (3, 0.99), (10, 0.75) \\
        &$\mu$ prior: $k$ value for $\sigma_\mu$ &0.5, 1, 5\\
        & Poisson rate for \# centers: $\lambda_c$ & 5, 25\\
    \hline
  Hetero-AddiVortes 
        & Sigma prior: $(\nu,q)$ & (3, 0.90), (3, 0.99), (10, 0.75) \\
        &$\mu$ prior: $k$ value for $\sigma_\mu$ &0.5, 1, 5\\
        & Poisson rate for \# centers: $\lambda_c$ & 5, 25\\
        
    \hline

    BART & Sigma prior: $(\nu, q)$ combinations &(3, 0.90), (3, 0.99), (10, 0.75)\\
        &$\mu$ prior: $k$ value for $\sigma_\mu$ &0.25, 0.5, 1, 2, 5, 10\\
        \hline
    HBART 
    &Sigma prior: $(\nu, q)$ combinations &(3, 0.90), (3, 0.99), (10, 0.75)\\
        &$\mu$ prior: $k$ value for $\sigma_\mu$ &0.25, 0.5, 1, 2, 5, 10\\
        \hline

  \end{tabular}
  \caption{A table showing the cross-validation values for competing methods.}
  \label{tab:CVvalues}
\end{table}

The box plots reveal that the RMSE values for all methods are comparable, indicating similar levels of accuracy in point predictions. However, heteroscedastic AddiVortes outperforms the competitors in terms of the e-statistic, which measures the similarity between predicted and true distributions. This substantial improvement in the e-statistic suggests that AddiVortes is not only capturing the central tendency of the data but also providing a better overall representation of the predictive distribution, particularly in scenarios with non-constant variance. This demonstrates its ability to model uncertainty and variability more effectively, offering a significant advantage in applications where understanding the full distribution is critical.

\begin{table}[ht]
  \centering
  \begin{tabular}{llll}
    \hline
    
    \textbf{Method} & e-statistic & RMSE \\
    & $\bm{(50\%,75\%)}$ & $\bm{(50\%,75\%)}$ \\ 
    \hline 
     Heteroscedastic AddiVortes & (0.3223,~0.4095) &(4644.7,~4926.2) \\ 
     Homoscedastic AddiVortes & (0.8492,~1.2699) &(4704.7,~5006.8) \\ 
     HBART & (0.3670,~0.8278) &(5140.7,~5390.0) \\
     BART & (0.8493,~1.2699) &(4693.7,~4842.8) \\
    \hline    
  \end{tabular}
  \caption{(50\%, 75\%) quantiles of e-statistics and RMSE values for each method.}
  \label{tab:quantiles}
\end{table}

\section{Million Songs Dataset}\label{million_songs_dataset}

To evaluate the heteroscedastic AddiVortes model in a more complex setting, we applied it to the year prediction task from the Million Songs Dataset (MSD). This dataset comprises 515,345 observations spanning 28,223 artists from 1922 to 2011, each with 90 predictor variables that encode audio features such as timbre averages (12 predictors) and timbre covariances (78 predictors). Predicting a song's release year from its audio features poses a challenging regression problem in a large \(n, p\) setting, making it an excellent test case for our model.

Previous studies, including those by \cite{Million_Songs}, explored prediction models for this dataset, such as \(k\)-nearest neighbors and the Vowpal Wabbit algorithm. We compared standard main-effects linear regression, AddiVortes, and heteroscedastic AddiVortes models, using default priors for the Bayesian approaches. Table \ref{tab:RMSE_Million_songs} summarizes the Root Mean Squared Error (RMSE) for year prediction on the test set, showing that the linear and \(k\)-NN models performed worse than the AddiVortes and heteroscedastic AddiVortes models. Notably, both AddiVortes and heteroscedastic AddiVortes demonstrated competitive predictive accuracy, with the latter excelling due to its ability to model heteroscedasticity.

\begin{table}[ht]
  \centering
  \begin{tabular}{llllllll}
    \hline
    
    Model & H-AV & AV & HBART  & Linear & 1-NN & 50-NN & Vowpal \\
    &&&&Regression&&&Wabbit \\ 
    \hline 
     RMSE & 8.97 & 9.10 & 9.08 & 9.51 & 13.99 & 10.20 & 8.76 \\ 
    \hline    
  \end{tabular}
  \caption{The RMSE values of all the competing methods for the test set of the million songs dataset.}
  \label{tab:RMSE_Million_songs}
\end{table}

Convergence diagnostics, including trace plots of the variance \(s^2(\bm{x})\) and posterior samples of \(\sigma\), confirmed reasonable convergence for both models. The heteroscedastic AddiVortes model revealed substantial heteroscedasticity within the dataset, as evidenced by the posterior distribution of \(s^2(\bm{x})\), which ranged from 1.04 to 51.9, compared to a constant posterior mean of 8.8 for \(\sigma\) in the AddiVortes model. The dataset's inherent imbalance, with most songs released post-1980 and relatively few observations from the early decades, does not lead to a different variance process, but the sparsity leads to more uncertainty. 

  \begin{figure}[H]
        \centering
        \includegraphics{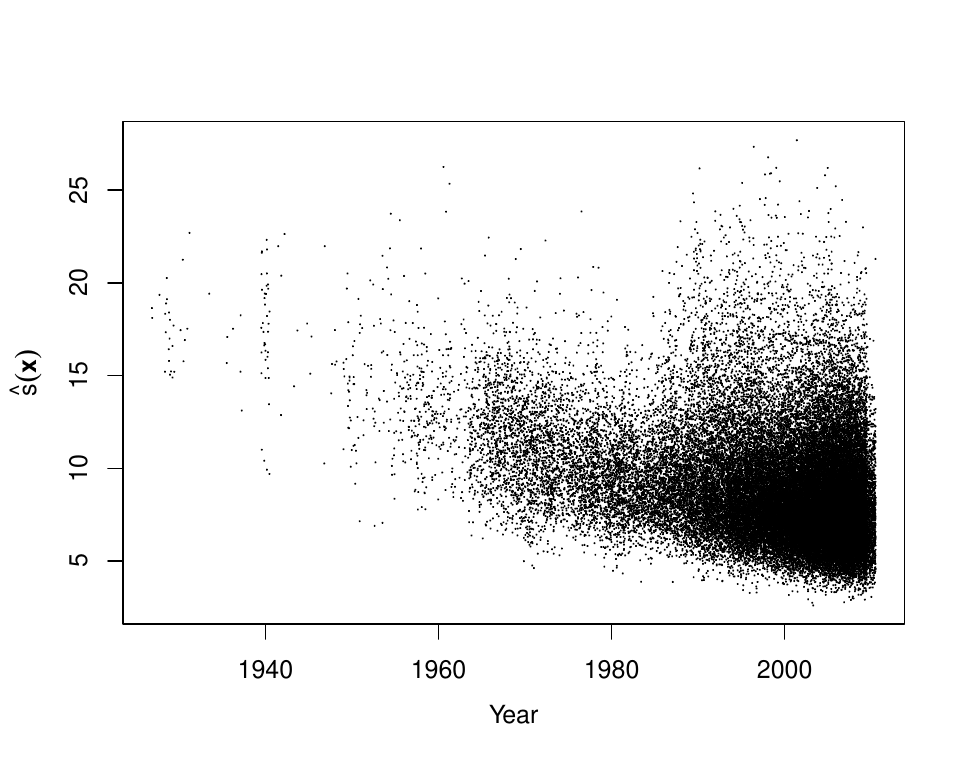}
        \caption{Plot of \(\hat{s}(x)\) versus year for the million songs test set. Note the values of year have been perturbed to more clearly convey how sample size varies with year}
        \label{fig:Year_vs_sd}
    \end{figure}

An analysis of the posterior average standard deviation \(\hat{s}(\bm{x})\) indexed by year showed that high predictive uncertainty is not solely due to sparse data in early years; heteroscedastic AddiVortes identified regions of high uncertainty even in recent years with abundant data. This highlights the model's ability to differentiate between uncertainty driven by data sparsity and intrinsic complexity in the predictor-response relationship. Thus, heteroscedastic AddiVortes serves as a valuable tool for exploring complex data patterns, pinpointing areas where predictive uncertainty is inherent and guiding further model refinement.

\section{Conclusion}\label{conclusion}

This paper introduced the heteroscedastic AddiVortes model, an extension of the Bayesian additive Voronoi tessellations framework, designed to simultaneously model both the conditional mean and variance of a response variable. By employing a sum-of-tessellations approach for the mean and a product-of-tessellations approach for the variance, the model offers a flexible, nonparametric way to capture complex relationships and heteroscedastic patterns in data.

The heteroscedastic AddiVortes model's key innovation lies in its ability to account for predictor-dependent variability, providing richer inferences than traditional homoscedastic models. The additive representation of the mean function enables intuitive modeling of main effects and interactions, while the multiplicative variance structure allows for precise, localized adjustments to the dispersion, reflecting changes in scale driven by specific predictors. This dual-layer approach ensures that both central tendency and variability are accurately modeled, offering a comprehensive view of the predictive distribution.

Empirical results demonstrate the model's effectiveness in capturing complex data patterns across diverse applications, ranging from simulated examples to real-world datasets. The ability to adaptively identify regions of heteroscedasticity and to provide interpretable insights into the sources of variation makes the heteroscedastic AddiVortes model a powerful tool for data exploration and prediction. Moreover, the use of conjugate priors and efficient MCMC sampling facilitates practical implementation, even in high-dimensional settings.

The heteroscedastic AddiVortes model represents a significant step toward flexible, interpretable Bayesian regression, empowering practitioners to make more informed and accurate inferences about their data.

	\section{Conflict of Interest statement}
	
	The authors report there are no competing interests to declare.
	\newpage
	\bigskip
	\begin{center}
		{\large\bf SUPPLEMENTARY MATERIAL}
	\end{center}
	
	\begin{description}
		
		\item[Online Appendices:] provides additional details of the default hyperparameter selection; cross-validation ranges used in our analysis and adjustments to the algorithm. (Online Appendices.pdf, pdf file)
		
		\item[GitHub Respiratory:] a repository on GitHub with all the code and data sets to run the algorithm and produce all the figures in this paper.\bf{ \href{https://github.com/Adam-Stone2/AddiVortes}{AddiVortes R code and data sets}.}
		
	\end{description}
	
	\bibliography{ref}
	
\end{document}